\documentclass[useAMS,usenatbib,fleqn]{mn2e}
\usepackage{amsmath,times,psfig}

\newcommand{\ale}{\ \raisebox{-.3ex}{$\stackrel{<}{\scriptstyle \sim}$}\ }
\newcommand{\age}{\ \raisebox{-.3ex}{$\stackrel{>}{\scriptstyle \sim}$}\ }

\title[The decline of GRS1915$+$105]{The decline and fall of GRS1915$+$105: 
the end is nigh?}
\author[M. R. Truss and C. Done]{Michael Truss$^{1,2}$\thanks{m.r.truss@durham.ac.uk} and Chris Done$^{1}$\\
$^{1}$Department of Physics, University of Durham, South Road, Durham, DH1 3LE\\
$^{2}$School of Physics \& Astronomy, University of St Andrews, North Haugh, St Andrews, Fife, KY16 9SS, UK}

\begin{document}

\date{Accepted ~~~ Received ~~~}

\pagerange{\pageref{firstpage}--\pageref{lastpage}} \pubyear{2005}
\maketitle

\label{firstpage}

\begin{abstract}
The galactic microquasar GRS1915$+$105 has been in a continuous state
of outburst since 1992, over 20 times longer than any other black hole
X-ray transient. Assuming that the outburst is powered via accretion
of an irradiated gaseous disc, we calculate how the predicted outburst duration 
varies according to the efficiency of the self-irradiation mechanism. At least 
one current model leads to the conclusion that the end of the outburst is
imminent. The timing of the decline of GRS1915$+$105, whenever it arrives,
will be an excellent discriminator of the self-irradiation mechanism
in X-ray transients, allowing us to infer the fraction of the disc that is
heated by the incident X-rays and the magnitude of the mass loss rate in the form of a wind.

\end{abstract}


\section{Introduction}
The galactic microquasar GRS1915$+$105 (V1487 Aql) lay undiscovered in
quiescence until 1992, when it was identified as an extremely bright X-ray 
transient \citep{cas}. The outburst that started in that year has continued to the 
present day, with no sign of an imminent decline. The binary comprises a 
$14 \pm 4.4 \, \rm M_{\odot}$ black hole accreting from a companion of about a solar mass 
\citep{har}, and remains the brightest accreting black hole in the galaxy, spending
much of its time at a super-Eddington X-ray luminosity \citep{don}.

The unusually long outburst is over 20 times longer than any other black hole 
X-ray transient outburst. The duration is linked to the size of the accretion disc, 
which is very large: GRS1915$+$105 has an extremely long orbital period of 
33.5 days, and as such the disc will have a radius of several $10^{12} \, \rm 
cm$. However, the reservoir of mass in the disc that is available to fuel an 
outburst is finite, and in this work we use a simple calculation to show how long
the outburst is likely to continue if the current mean accretion rate is
maintained. We present this calculation in Sections 2 and 3 below. In Sections 4 and 5,
we discuss the implications for our understanding of the accretion process in
X-ray transients. 

\section{Fuelling the outburst}
Assuming that the outburst of GRS1915$+$105 is fuelled by the accretion of gas
contained in a disc, an absolute upper limit for the outburst duration can be 
found by considering the time taken to accrete the entire reservoir of mass. 
\citet{don} made a simple estimate for the outburst duration of 10 years, based 
on an outer disc radius $R_{\rm disc} \sim 10^{12} {\rm cm}$. Clearly, this 
duration has been exceeded and this is the motivation for undertaking a more
detailed calculation. We begin by considering the maximum available disc mass. 
The outburst duration will be given by
\begin{equation}
t_{\rm max} = \frac{M_{\rm disc}}{\langle \dot M_{\rm disc} \rangle},
\label{time}
\end{equation} 
where $M_{\rm disc}$ is the mass of the accretion disc at the beginning of the 
outburst. We assume that the mass transfer rate from the donor star, 
$-\dot M_2$, remains constant throughout the outburst and consider a 
time-averaged central accretion rate onto the black hole, $\langle \dot M_1 
\rangle$ and a time-averaged wind mass loss rate $\langle \dot M_{\rm wind} 
\rangle$. The time-averaged rate of mass loss from the accretion disc is then
\begin{equation}
\langle \dot M_{\rm disc} \rangle = \langle \dot M_1 \rangle + \langle \dot 
M_{\rm wind} \rangle - \dot M_2.
\label{mdot}
\end{equation} 
GRS1915$+$105 spends much of its time radiating with a super-Eddington luminosity \citep{don}, so we infer that the mass accretion rate is consistently very high. Assuming a distance $d = 12.5 \,\rm kpc$ (\citet{gre}, see the discussion in Section 4), the mean 
luminosity is close to Eddington \citep[Figure 5]{don}, implying a mass accretion rate at 
the black hole of order $\dot M_{\rm Edd} \sim 2 \times 10^{19} \, \rm gs^{-1}$. This is much 
larger than the estimate for the mass transfer rate from the companion, even taking into 
account its evolved nature. We use the formula for mass transfer driven by
nuclear evolution given in equation 6 of \citet{kfkr}, which for the parameters 
of GRS1915$+$105 and a core mass of $0.28 \, \rm M_\odot$ \citep{vil}
gives $-\dot M_2 \simeq 10^{-8} \, {\rm M_\odot yr^{-1}} = 6.3 \times 10^{17} \, {\rm gs^{-1}}$.

We now wish to estimate the mass of the disc at the start of the outburst. Ideally, this would
be determined from the duration of an 
interval of quiescence, but this is not possible with GRS1915$+$105. Only one 
outburst has ever been observed - the current one - and the quiescent interval 
for such a large disc is likely to be centuries. Therefore, we must assume an 
appropriate surface density profile for the disc, $\Sigma(R)$, at the onset of an 
outburst.

The mass of the disc is given by
\begin{equation}
M_{\rm disc} = \int_0^{R_{\rm disc}} 2\pi R \Sigma(R) \,{\rm d}R,
\label{mass0}
\end{equation}
and we assume that the surface density at all radii in the disc is equal to the 
critical surface density required to trigger an outburst via the thermal-viscous 
instability:
\begin{equation}
\Sigma = \Sigma_{\rm max} = 11.4 \alpha_{\rm c}^{-0.86} M_{\rm 1}^{-0.35} 
R_{\rm 10}^{1.05}
\label{mass1}
\end{equation}
\citep{can}, where $R_{10}$ is the radius in units of $10^{10} \, \rm cm$. We take the masses $M_1 = 14 \, \rm M_\odot$, $M_2 = 1 \, \rm M_\odot$ 
and a cool state viscosity $\alpha_{\rm c} = 0.02$. We discuss the effects of 
taking different values for these parameters in Section 4. Integrating equation \ref{mass0}, 
we are left with
\begin{equation}
M_{\rm disc}=3.4 \times 10^{28}\left(\frac{\alpha_{\rm c}}{0.02}\right)^{-0.86}\left(\frac{M_1}{14}\right)^{-0.35}R_{\rm disc,12}^{3.05} \,\,\,\,\,\rm g
\label{mass3}
\end{equation}
where the radius is now scaled to units of  $10^{12} \, \rm cm$. 

It is not immediately clear how to make a secure estimate for the maximum 
outer radius of the disc in quiescence.  The maximum possible streamline radius 
in the three-body model of \citet{pac} is about $0.5a$. For $P_{\rm orb} = 33.5 \, 
\rm d$ and $M_1 + M_2 = 15\,M_{\odot}$, the binary separation $a = 7.5 \times 
10^{12} \, \rm cm$ so this estimate for the radius gives $R_{\rm disc} \simeq 3.7
\times 10^{12} \, \rm cm$. However, this is likely to be an overestimate. Taking 
a cue from low mass ratio cataclysmic variables, the radius of the disc in 
quiescence is always much less than the maximum streamline radius, only 
approaching it as the disc expands in the hot, highly eccentric outburst state. A 
more realistic estimate for the disc radius in a quiescent X-ray transient is given
by \citet{shah}, who use angular momentum conservation arguments to show
that in the case of negligible accretion onto the primary, $R_{\rm disc} = 
1.36R_{\rm circ}$.  An expression for the circularisation radius $R_{\rm circ}$ is
given by
\begin{equation}
\frac{R_{\rm circ}}{a} = 0.0859q^{-0.426}
\end{equation}
\citep{hes}, which is accurate for $0.05 \le q < 1$. For $q={1 \over 14}$, we have 
$R_{\rm circ} = 0.26a$, giving us the refined estimate $R_{\rm disc} = 2.7 
\times 10^{12} \, \rm cm$. 

For $R_{\rm disc,12} = 2.7$, equation \ref{mass3} gives an estimate for the disc
mass $M_{\rm disc} = 7.0 \times 10^{29} \,\rm g$. It follows from equations \ref{time} 
and \ref{mdot} that the time taken to accrete the entire disc is in fact extremely long: $t_{\rm max} \sim 1150$ years if there is no mass loss due to a wind. The huge discrepancy between this  estimate and that in \citet{don} is due to differences in our estimates of $R_{\rm disc}$ and $M_{\rm disc}$. Clearly, the outburst duration is sensitive to $R_{\rm disc}$ 
and the initial surface density profile $\Sigma(R)$: in the simple model described 
above, the disc mass scales as $R_{\rm disc}^{3.05}$. In the next section, we 
make a more detailed estimate of the outburst duration by considering a more 
realistic surface density profile and estimating the fraction of the total disc 
mass that is available to be accreted onto the black hole.

\section{Outburst duration}

The simple calculation described above makes two important assumptions.  The 
first assumption is that the surface density profile follows $\Sigma(R) = 
\Sigma_{\rm max}$ at all radii at the onset of the outburst. This is not physically 
realistic, as the only requirement to trigger an outburst is that $\Sigma(R) > 
\Sigma_{\rm max}$ at a single radius. The second is that the entire disc mass is
consumed in an outburst. 

We begin by addressing the problem of the surface density profile at the onset 
of an outburst. Detailed models of outburst cycles in X-ray transients 
\citep{dub} show that the surface density crosses the $\Sigma_{\rm max}$ 
threshold in the {\em inner} region of the disc. Indeed, the model presented in 
Figure 15 of \citet{dub} shows that $\Sigma$ follows $\Sigma_{\rm max}$ 
closely only in the inner 10 - 15 \% of the disc and flattens off somewhat at 
larger radii. If the surface density profile of the disc in GRS1915$+$105 follows 
a similar structure, we calculate the total disc mass to be much smaller than the
value given in Section 2 above. Very simply, assuming that 
$\Sigma=\Sigma_{\rm max}$ for $0 \le R \le 0.1R_{\rm disc}$ and
$\Sigma=\Sigma_{\rm max}(R=0.1R_{\rm disc})$ for $0.1R_{\rm disc} \le R \le 
R_{\rm disc}$, we have
\begin{equation}
M_{\rm disc} = M(R<0.1R_{\rm disc}) + M(R>0.1R_{\rm disc}).
\end{equation}
In practise, the first term is negligible, and to a very good approximation
\begin{equation}
M_{\rm disc} = \int_{0.1R_{\rm disc}}^{R_{\rm disc}} 2\pi R \Sigma_{\rm max}(0.1R_{\rm disc}){\rm d}R,
\end{equation}
or
\begin{equation}
M_{\rm disc} = 1.3 \times 10^{28} \left(\frac{\alpha_{\rm c}}{0.02}\right)^{-0.86} \left(\frac{M_1}{14}\right)^{-0.35} R_{\rm disc,12}^2.
\label{mass2}
\end{equation} 
Note that the coefficient in equation \ref{mass2} is specific to this system, because it assumes a value for $R_{\rm disc}$ that depends on the binary parameters. With $R_{\rm disc,12} = 2.7$, $\alpha_{\rm c}=0.02$ and $M_1=14$, this gives $M_{\rm disc}=9.5 \times 10^{28} \,\rm g$, reducing 
the maximum outburst time to about 160 years (again assuming zero mass loss 
in a wind).

However, this is the maximum mass of the cold disc at the start of the outburst,
 but in an accretion disc as large as the one in GRS1915$+$105, a significant 
fraction of the outer parts of the disc may be too cool to support an outburst at 
all \citep{ham}. The only way that a significant fraction of such a large disc can 
remain in the hot, ionised state is by self-irradiation.  Heating by incident X-ray
radiation produced near the accretor (or scattering of some small fraction of the 
radiation back down onto the disc by some form of corona) prevents the disc 
from switching back into the cool state, so prolonging the outburst \citep{dub}. 
The radius of influence of the incident X-rays, $R_{\rm irr}$  is usually estimated 
by matching the irradiation temperature to the hydrogen ionization temperature 
such that
\begin{equation}
T_{\rm irr}^4 = T_{\rm H}^4 =\epsilon \frac{L_{\rm X}}{4\pi\sigma 
R_{\rm irr}^2} = \frac{\epsilon \eta \dot Mc^2}{4\pi\sigma R_{\rm irr}^2}
\label{tirr}
\end{equation}
where the constant of proportionality $\epsilon$ depends on the geometry of the disc, the 
nature of the illuminating X-ray source and the X-ray albedo of the gas. We will refer to $\epsilon$ as the {\em irradiation efficiency} to distinguish it from the {\em accretion efficiency}, $\eta$. This notation differs slightly from that used by \citet{dub}, who use $C$ for the irradiation efficiency. \citet{kkb} and \citet{kin00} have pointed out that the Eddington limit for accretion imposes a 
limit on $R_{\rm irr}$, leading to the conclusion that systems with orbital periods longer than 
about 2 days must be transient, because at these periods $R_{\rm irr}$ can 
never be larger than $R_{\rm disc}$, even for accretion beyond the Eddington 
limit. The parameters used in this model give
\begin{equation}
R_{\rm irr} = 2.3 \times 10^{11} \left( \frac{\eta}{0.1} \right)^{\frac{1}{2}} 
\dot M_{18}^{\frac{1}{2}} \,\, \rm cm
\label{rirr}
\end{equation}
\citep{kin98,tru04}, where $\dot M_{18}$ is the central accretion rate in units of 
$10^{18} \rm gs^{-1}$. For accretion at the Eddington limit, with efficiency 
$\eta = 0.1$, this predicts
\begin{equation}
R_{\rm Edd} \simeq 10^{12} \,\rm cm.
\label{redd1}
\end{equation}

Other models - with an equally sound observational footing - predict a
more efficient irradiating flux than the one assumed by \citet{kin98}. \citet{dub99} and \citet{dub} use a constant of 
proportionality that is typically about seven times larger in equation \ref{tirr}, 
which leads to an estimate for $R_{\rm Edd}$ more than twice as large. Indeed, 
the alternative spherical inner X-ray source geometry considered by 
\citet{kin98} also leads to a larger estimate for $R_{\rm Edd}$. 
If we make no assumptions about the irradiation efficiency, we can write the more general expression
\begin{equation}
R_{\rm irr} = 2.7 \times 10^{11} \left( \frac{\epsilon}{10^{-3}} \right)^{\frac{1}{2}} 
\left(\frac{\eta}{0.1}\right)^{\frac{1}{2}} \dot M_{18}^{\frac{1}{2}} \,\, \rm cm.
\label{rirr2}
\end{equation}
We will return to 
this point in more detail in Section 4, but for the time being we use equations \ref{rirr} and 
\ref{redd1} as our example, because they give the smallest value of 
$R_{\rm Edd}$ and hence leads to the shortest possible predicted outburst 
duration.

Since the mean source luminosity of GRS1915$+$105 is observed to be around 
Eddington, $R_{\rm Edd}$ represents the maximum radius of the hot, outburst 
region of the disc. The remainder of the disc outside this point stays too cool to 
participate in an outburst, so the fraction of total disc mass accreted is much 
less than unity for a large disc \citep{shah}. Adding this piece of information to 
the more realistic estimate of the surface density profile allows us to calculate 
the maximum available mass of gas for the outburst. Repeating the calculation 
described above, but now using a maximum radius of $10^{12} \,\rm cm$ gives 
$M_{\rm max} = 1.2 \times 10^{28}\,\rm g$. This is simply the disc mass inside 
$10^{12} \,\rm cm$.
\begin{table}
\begin{tabular}{|c|c|c|c|c|} \hline
$R_{\rm irr} \times 10^{12} \,\rm cm$ & $\epsilon \times 10^{-3}$ & $M \times 10^{28} \,\rm 
g$ & $t_{0} \, \rm (yr)$ & $t_{\rm wind} \, \rm (yr)$ \\ \hline
0.5 & 0.17 & 0.29 & 4.7 & 2.3 \\
1.0 & 0.69 & 1.3 & 21 & 10 \\
1.5 & 1.6 & 2.9 & 47 & 23 \\
2.0 & 2.8 & 5.2 & 85 & 42 \\ 
2.5 & 4.3 & 8.2 & 130 & 66 \\ 
2.7 & 5.1 & 9.5 & 160 & 76 \\ \hline
\end{tabular}
\caption{Predicted maximum outburst durations for various irradiated fractions of
the accretion disc. The calculation of the available disc mass is described in 
Section 3 and the durations assume that all of the mass inside $R_{\rm irr}$ is 
accreted during the outburst. The irradiation efficiency, $\epsilon$, is calculated from 
equation \ref{rirr2}. Columns 4 and 5 give the predicted maximum
duration assuming $\langle \dot M_{\rm wind} \rangle = 0$ and $2 \times 10^{19} 
\, \rm gs^{-1}$ respectively.}
\end{table}
If the mean accretion rate continues at its current value (which is 
approximately $2 \times 10^{19} \, \rm gs^{-1}$) and 100\% of the mass 
originally inside $R = 10^{12} \,\rm cm$ is accreted, we expect the outburst to 
last approximately 20 years if there is no mass lost in the form of a jet or a wind.

However, there is considerable evidence for mass loss in this system via 
a wind. Relativistic velocities which might be appropriate for a jet mean that its mass loss rate 
can be small compared to the mass accretion rate, even if it makes a significant
contribution to the energy budget \citep{nay}. The same is {\em not} true for a much slower
outflow such as a wind. Numerical simulations by \citet{proga} show that 
accretion in Galactic binary systems with high Eddington fractions can power a
strong disc wind. These are driven by radiation pressure on the electrons as 
opposed to line driven as the material is so highly ionised it has little absorption 
opacity. At Eddington, the mass loss rate in this wind should be comparable to 
the mass accretion rate \citep{proga}, and there is observational evidence for 
such high mass loss rates in GRS1915$+$105 from detection of blueshifted, 
extremely ionised X-ray absorption lines \citep{lee}. If approximately 
$\dot M_{\rm Edd}$ is being lost to the wind, then the outburst timescales need to 
be halved. Table 1 shows the predicted maximum outburst durations for a range
of irradiated disc fractions, with and without a significant wind mass loss rate. We
interpret these two durations (no wind and an Eddington-rate wind) as reasonable 
upper and lower limit estimates for the outburst time-scales for each given irradiated
fraction of disc. If an Eddington wind loss rate is taken into account, we can see immediately that since the outburst has already progressed for at least 13 years, this supports the assertion made by \citet{kin98} that the more appropriate source geometry for a black hole system in outburst is in fact that of a central point source, leading to slightly stronger irradiation and a hot area of disc beyond $10^{12}\,\rm cm$. Furthermore, it is clear that for $R_{\rm irr} \ale 1.5 \times 10^{12} \,\rm cm$, the total mass of irradiated gas will be consumed in the next few years and we would expect the outburst to terminate.

\section{Discussion}

We have shown that the total mass of the accretion disc in GRS1915$+$105 just 
before the onset of an outburst is of order $\sim 10^{29} \,\rm g$. At the current 
mean mass accretion rate, inferred for an accretion efficiency $\eta \sim 0.1$, 
this is enough to power the outburst for 160 years. However, given the large 
scale of the disc, we surmise that a large fraction of the outer regions will
remain too cool to sustain an outburst. Thus even the mass added to the outer edge of the disc from the companion star cannot replenish the hot inner disc region. Instead, it is stalled
at larger radii where it does not participate in the outburst. In this scenario, equation \ref{mdot}
only involves the mass accretion rate and the wind loss rate. This is important, because the
outburst time-scale is not affected by uncertainties in the mass transfer rate, which is extremely sensitive to the secondary core mass \citep{kfkr,rit}. 

In fact, the values in Table 1 are calculated including $-\dot M_2 = 10^{-8} \rm M_\odot yr^{-1}$ in equation \ref{mdot}, but since $-\dot M_2 << \dot M_1$, this is no different from the case
$-\dot M_2 = 0$. If  $-\dot M_2$ is higher than our estimate, either due to uncertainties in the evolutionary state of the system or due to an irradiation-induced
burst of mass-transfer from the companion star, the outburst still cannot be prolonged because the additional mass remains in the cool outer disc. We can estimate the required $-\dot M_2$ at which this assumption breaks down. \citet{dub} give an expression for the mass transfer rate required to trigger
a heating wave at the outer edge of the disc:
\begin{multline}
\dot M = 3.3 \times 10^{16} \, \delta^{-0.5} \left(\frac{\alpha}{0.02}\right)^{0.2}\left(\frac{M_1}{7}\right)^{-0.9}
\\
\left(\frac{T_{\rm c}}{2000 \,\rm K}\right) R_{\rm disc,11}^{2.6} \,\,\, \rm gs^{-1}
\end{multline}
where $\delta$ is a parameter with typical value $0.05 - 0.1$. So, for GRS1915$+$105 at $R_{\rm disc,11} = 27$, $\delta=0.1$ and $T_{\rm c} = 1000 \rm K$, we have a required mass transfer rate $\dot M = 1.5 \times 10^{20} \,\rm gs^{-1} \sim 2 \times 10^{-6} \rm M_\odot yr^{-1}$. While this is below the rate  required to make the source persistent, it is still more than two orders of magnitude larger than would be
expected from standard theories of binary evolution.

It is instructive to quantify the potential effect on our calculations of uncertainties in the 
accretion process itself and in the observed parameters of GRS1915$+$105. Taking the 
simplest case where $\dot M_{\rm wind} = 0$ and $-\dot M_2 = 0$, we see from equations
\ref{time} and \ref{mdot} that $t_{\rm max} \propto {M_{\rm disc} / \dot M_1}$. Using equations \ref{mass2} and \ref{tirr}, we find that
\begin{equation}
t_{\rm max} \propto M_1^{-0.35} \alpha_{\rm c}^{-0.86} \eta \epsilon,
\label{prop}
\end{equation}
allowing us to immediately identify the relative importance of uncertainties in the different parameters. It is surprising - but nevertheless a very desirable aspect of the model - that neither the luminosity nor the inferred accretion rate enter this relation at all. A lower accretion rate leads to a lower $R_{\rm irr}$ 
and a smaller available mass of hot gas. This means that the effect that any uncertainties in the distance have on the outburst time-scale is weakened. In fact, in our simple model for the surface density of the 
disc, where the profile is flat at most radii, we find in equation \ref{mass2} that the mass scales as $R^2$. In this case, since $R_{\rm irr} \propto L_{\rm X}^{\frac{1}{2}}$, the accretion rate doesn't 
appear in equation \ref{prop}. This is very important, because observational uncertainties in the distance - here we use $d = 12.5 \rm kpc$ after \citet{mir} but a more recent work using proper motions of jet components places the source about $2 \,\rm kpc$ closer \citep{mil} - do not make any difference at all to the predicted outburst duration.

The black hole mass enters equation \ref{prop} rather weakly, though the observational uncertainties in this quantity are large. If the system is aligned with the jet,
\citet{har} state that $M_1=14 \pm 4.4 \, \rm M_{\rm \odot}$ for a jet inclination 
$i=66 \pm 2^{\circ}$ \citep{fen}. If the jets are misaligned with the plane of the disc and are 
precessing, the range of possible masses widens. \citet{har} consider a system inclination 
offset by $10^{\circ}$ to the jets. At the extremes of this range, the mass could 
be anything from $11.6 \pm 3.3 \, \, \rm M_{\rm \odot}$ to $16.9 \pm 5.9 \, \rm M_{\rm \odot}$. 
Over the entire range of possible masses $8 {\rm M_\odot} < M_1 < 23 {\rm M_\odot}$, our predicted outburst duration only changes by a maximum of 21\%.

The quiescent viscosity, $\alpha_{\rm c}$ is a much more significant uncertainty. While the origin of viscous shear is well-understood for hot, ionized gases in terms of the magneto-rotational instability, our understanding of viscous processes in a cool, neutral gas is very limited. Therefore, an appropriate value for $\alpha_{\rm c}$ is hard to estimate. Our choice of $\alpha_{\rm c} = 0.02$ is well-motivated by the disc instability model and observations of quiescent intervals in dwarf novae, although rather different values have been suggested.
\citet{mey} argue that $\alpha_{\rm c} = 0.05$ based on a study of the quiescent intervals of shorter period X-ray transients. This would {\em decrease} our predicted time scales in Table 1 by a factor of two. Similarly, the accretion efficiency $\eta$ could be higher than 0.1 given that the black hole is spinning. A higher efficiency means that a smaller accretion rate is required to power the same luminosity, and would result in a longer outburst. $\eta=0.1$ is appropriate for a black hole spin $a \simeq 0.7$; at $a=0.9$, $\eta \simeq 0.15$ is more appropriate, leading to an outburst duration 50\% longer.
 
The key parameter in determining the size of the mass reservoir to power the outburst is the irradiation efficiency, $\epsilon$. Of the disc irradiation models considered, the smallest fraction of irradiated disc is predicted by the inner disc source geometry described in \citet{kin98}. Here, even for accretion at the Eddington limit, only the parts of the disc inside 
$R = 10^{12}\,\rm cm$ are illuminated by the incident X-rays. Defining this region 
as the only part of the disc capable of supporting an outburst, we calculate that 
its mass, of order $\sim 10^{28}\,\rm g$, is only 
sufficient to power an outburst for approximately 20 years assuming no wind loss. More efficient irradiating geometries mean that more of the disc can be illuminated, so increasing the mass available to power the outburst and hence its duration.
 
The only uncertainty not present in equation \ref{prop} is the magnitude of the mass loss in a
wind, and we can see immediately from Table 1 that there is a factor of two difference in 
outburst duration between the case of zero and Eddington wind losses. The observed wind loss rate is substantial \citep{lee}, but this depends on the (unknown) opening angle of the wind. Numerical simulations suggest that this should be fairly large \citep{proga}, in which case the inferred mass loss rates are comparable to the accretion rates required to sustain the outburst. In this case, the reservoir of available disc mass will last only half 
as long as expected from accretion alone. 

\section{Conclusions}
We have calculated upper and lower limits for the outburst time scale of GRS1915$+$105 for 
different values of the irradiation efficiency (i.e. different irradiation geometries). These limits correspond to zero and Eddington wind mass loss rates respectively. The time scales in Table 1 are computed for reasonable values of black hole mass, disc viscosity and accretion efficiency. The sensitivity of our results to uncertainties in these properties are discussed in Section 4, in particular equation \ref{prop}. 

The crucial factor that remains to be discovered is how efficient the irradiation actually is. It is clear that the mass budget for GRS1915$+$105 already seems very tight given all the competing factors. We are faced with the very interesting possibility that the outburst 
will come to an end in the next few years. If the outburst continues for substantially 
longer then we would have to conclude that there are additional factors at work.

The outer disc is the only feasible additional mass 
source in the system, and one way to tap this is via the wind. Scattering in this material can enhance the illumination of the outer disc, and there is observational evidence for this 
effect inferred from a detailed consideration of the outburst characteristics of 
neutron stars and black holes \citep{dub99}. Indeed, scattering of this kind may be
the only way to irradiate the disc at all: many simulations of discs irradiated by a
central source show that the disc puffs up and self-shields itself from the X-rays
\citep{can95,dub99}, in contradiction with observations showing conclusively that
discs in these systems {\em are} irradiated.

The fraction of X-rays scattered 
onto the disc $C \sim \tau_{es}\Omega/2\pi$, where $\tau_{es}$ is the electron 
scattering optical depth and $\Omega/2\pi$ is the solid angle subtended by the 
material. The wind simulations of \citet{proga} show $\Omega/2\pi\sim 0.3-0.5$, 
while the column density measured in the ionised absorber in GRS1915$+$105 
implies $\tau_{es}\sim 0.01$ (similar optical depths are inferred for the accretion 
disc coronal sources) i.e. $C \sim 5\times 10^{-3}$. If we assume that all of this X-ray flux incident on the disc goes into heating the gas, then $C\sim \epsilon$, giving a heated disc radius $\age 2 \times 10^{12} \, \rm cm$. However, if only a fraction of the incident X-rays heat the gas, we would expect $\epsilon$ and $R_{\rm irr}$ to be smaller.

The presence of the wind can give rise to a interesting feedback. A high accretion rate can lead to
a strong wind that may be associated with more efficient irradiation. This will lead to a further increase in the central accretion rate. However, this cannot continue unchecked, because too strong a 
wind will deplete the mass in the inner disc region, and the accretion rate will fall. We will explore these ideas in a later paper.

Our calculation can be applied to the outbursts of shorter period X-ray transients. In many ways these
are far simpler, because we would expect the whole disc to become irradiated and participate in the 
outburst. For example, taking the parameters of the black hole systems A0620-003 and GS2000+25, which  both have orbital periods around 8 hours, we find that the total mass consumed during each outburst is about 55-70\% of the calculated initial total disc mass. This
assumes $\eta=0.1$, zero wind mass loss and an exponential decay in central accretion rate calculated from the data given in \citet{che}. However, we should point out that we do not expect this simple approach
to work in all cases: while in general it is true that longer period systems tend to radiate more energy 
(and thereby have a higher inferred accretion rate), this is not always the case. For example, Aql X-1, with an orbital period of about 19 hours, does not undergo rare long outbursts, but frequent very short ones.

It is clear that there is still much to be learnt about accretion disc structure and how
it is affected by irradiation in X-ray transients: for example, can the disc be directly
illuminated or are the X-rays scattered down onto the disc? How efficient is the
irradiating flux? If the current outburst of GRS1915$+$105 goes on and on, the conclusion that a large fraction of the disc is irradiated is
inescapable. Since GRS1915$+$105 has the longest orbital period of any known transient
(by some distance) and hence the largest disc, we may be faced with the possibility that
{\em all} discs in X-ray transients can be fully, or almost fully, irradiated. The corollary of 
this is that should the outburst terminate in the next few years, we will have the exciting
opportunity to determine the fraction of the disc that was irradiated and learn much 
about the efficiency of the process and the importance of mass loss in a wind.

\section*{Acknowledgments}
MRT acknowledges a PPARC postdoctoral fellowship and the support of a 
PPARC rolling grant at the University of St Andrews. We are particularly grateful to the
referee, whose constructive comments and insight improved the paper significantly. 
We would also like to acknowledge Ed Cackett and Keith Horne for helpful discussions.

\label{lastpage}

\end{document}